\documentclass{PoS}

\title{Towards an effective field theory for vector mesons}

\ShortTitle{Towards an effective field theory for vector mesons}

\author{\speaker{Stefan Leupold} and Carla Terschl\"usen%
\\
        Department of Physics and Astronomy, Uppsala University, Sweden\\
        E-mail: \email{stefan.leupold@physics.uu.se}}

\abstract{The assumption that vector mesons dominate the interactions of hadrons with electromagnetism (vector-meson dominance 
--- VMD) provides an important phenomenological concept. On the other hand, a clear microscopic derivation is still missing
and there are cases where VMD drastically fails, e.g.\ for the omega transition form factor. 
In principle, effective field theories with their systematic expansion and power counting
could provide a tool to assess the validity of VMD and more generally to describe the interactions of vector mesons at low
energies. Though the systematic development is still in an infant stage we present here a Lagrangian for light pseudoscalar 
and vector mesons which is inspired by ideas from effective field theories. The Lagrangian is used to calculate 
electromagnetic meson form factors. It turns out that one can reproduce both the successes of VMD concerning
the pion form factors and the deviations from VMD concerning the omega transition form factor.}

\FullConference{50th International Winter Meeting on Nuclear Physics\\
                 23-27 January 2012 \\
                 Bormio, Italy}

\begin{document}

\section{Introduction}

A good understanding of the interactions of hadrons with electromagnetism is a key ingredient for several areas of hadron and
particle physics. To name a few examples: 
a) The first indication that the proton has an intrinsic structure came from elastic electron-proton
scattering \cite{kendall}. Follow-up experiments for elastic and (deep) inelastic scattering revealed more and 
more of the proton structure, nowadays quantitatively encoded 
in structure functions, electromagnetic form factors, generalized parton distributions and so on. 
b) The change of the properties of hadrons once they are placed in a strongly interacting environment is an active field of
research in heavy-ion and neutron-star physics. Such in-medium properties of hadrons can be accessed in 
dilepton spectra \cite{LMM,CBM}. 
c) A promising candidate to search for physics beyond the standard model is the gyromagnetic ratio of the muon. 
At present the largest uncertainty on the theory side for the standard-model calculation is provided by the 
hadronic contribution to this gyromagnetic ratio 
\cite{Jegerlehner:2009ry}. 

So, what do we know about the interactions between hadrons and real or virtual photons? The neutral vector mesons which have
the same quantum numbers as the photon can couple directly to the photon. Indeed, these vector mesons are prominently seen in 
the corresponding cross sections and decay rates. This gave rise to the notion of vector-meson dominance (VMD) \cite{sakurai}.
In the following, we will concentrate on electromagnetic meson form factors. We start with a brief review how well such
form factors are described by VMD. 

\begin{figure}[h!] 
  \centering
  \includegraphics[keepaspectratio,width=0.5\textwidth]{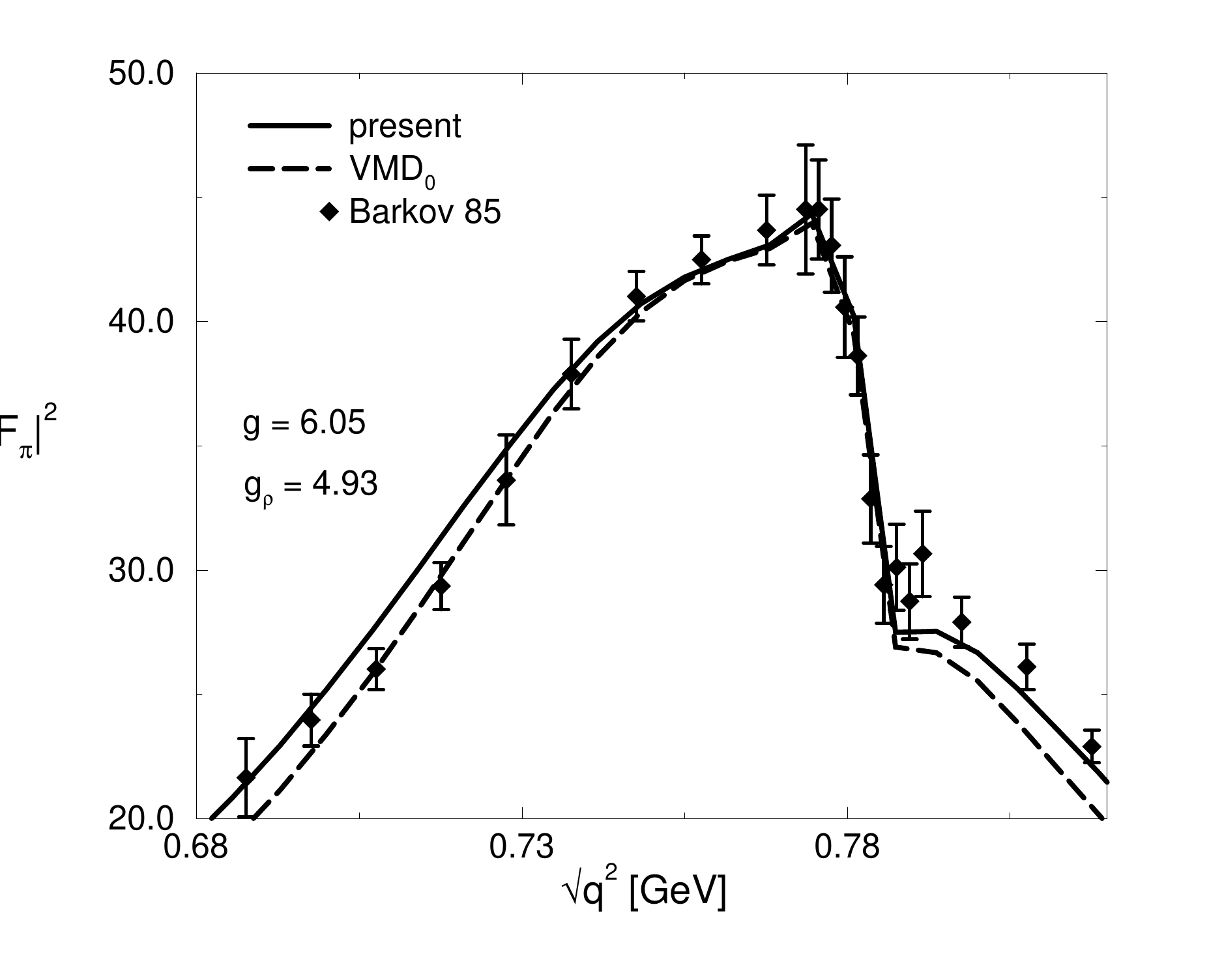} 
  \caption{The pion form factor as compared to VMD. Figure taken from \cite{Klingl:1996by}.}
  \label{fig:klingl}
\end{figure}

\begin{figure} 
  \centering
  \includegraphics[keepaspectratio,width=0.45\textwidth]{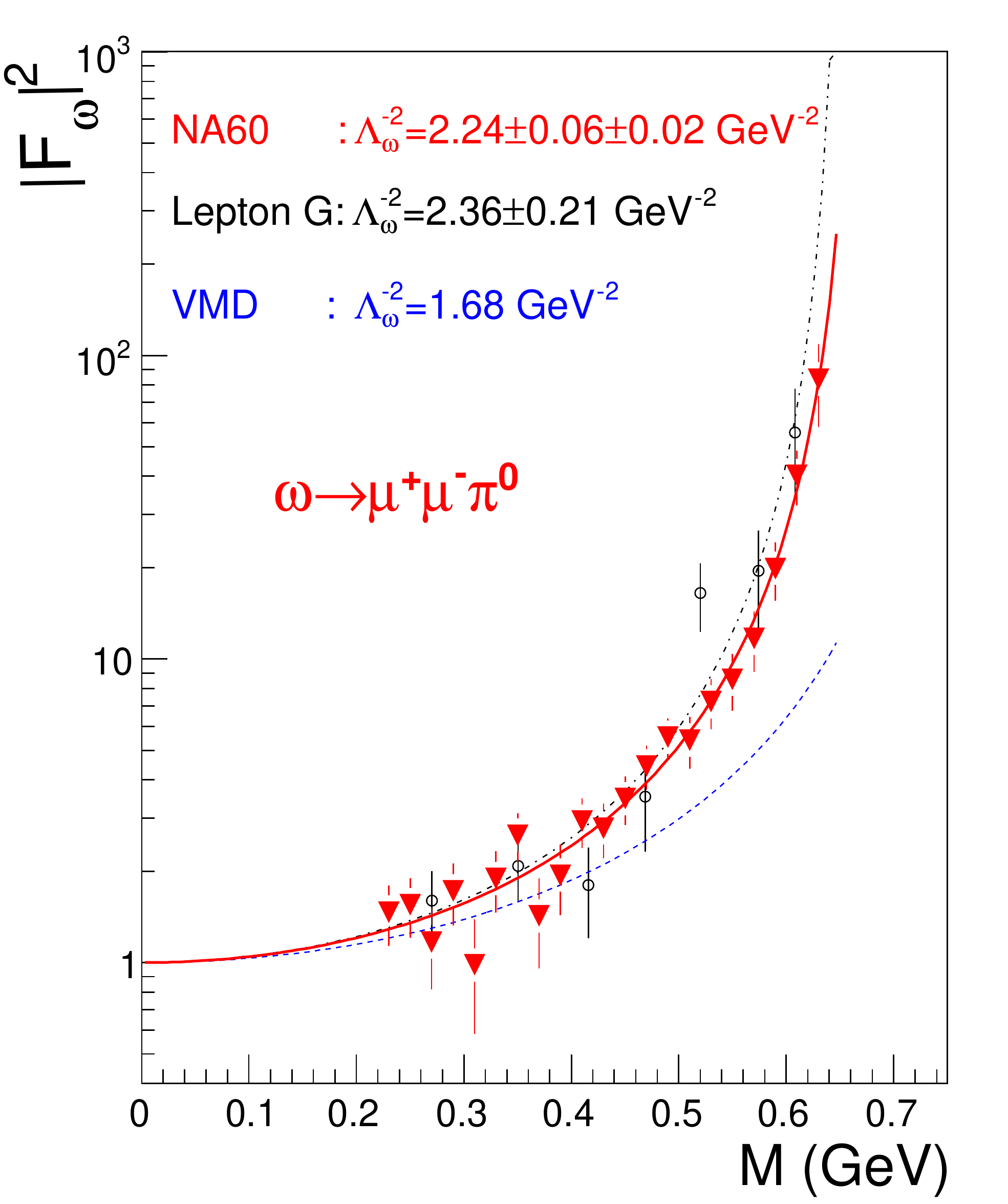} \hfill
  \includegraphics[keepaspectratio,width=0.45\textwidth]{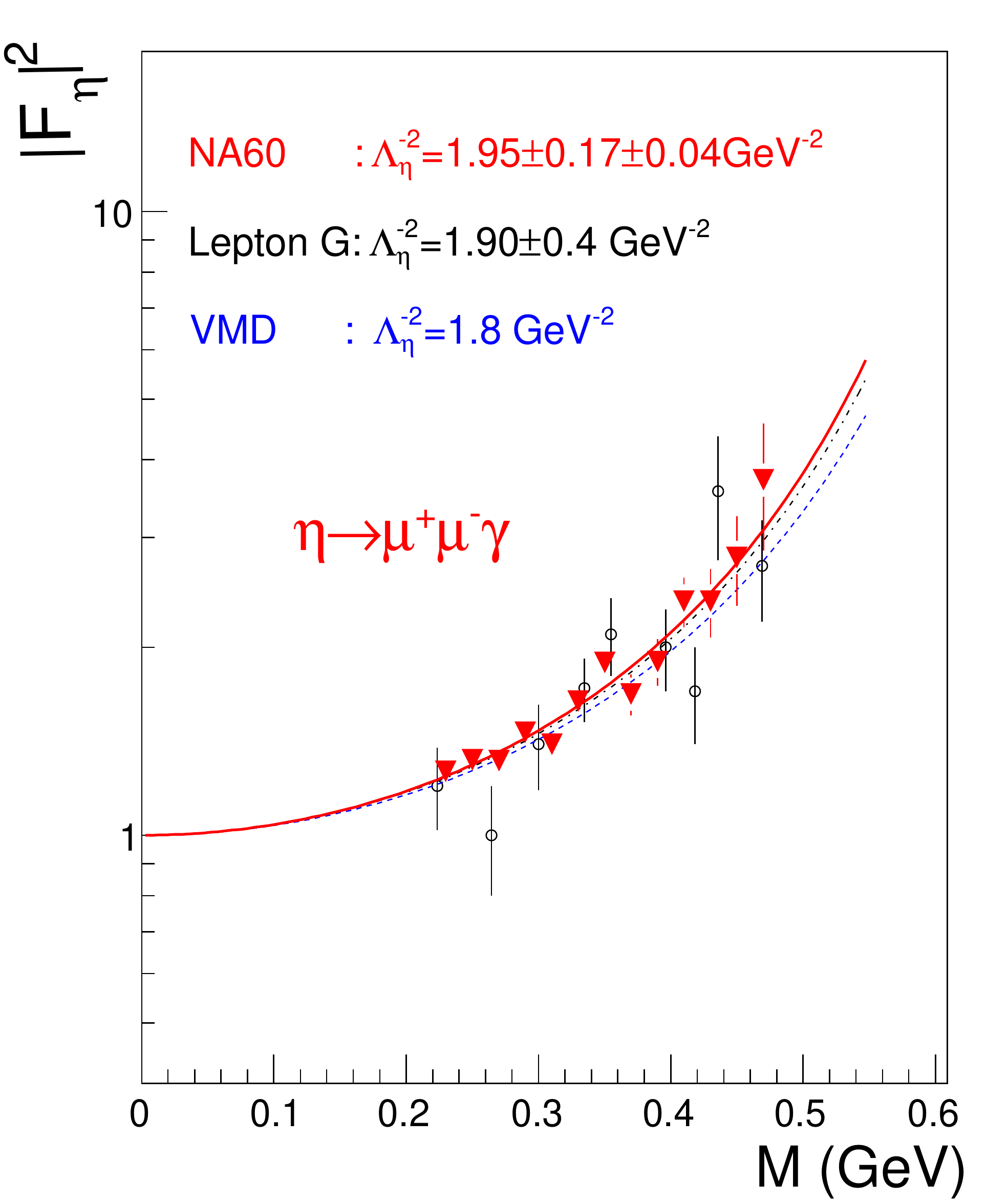}
  \caption{The transition form factors of omega to pion (left) and of eta to photon (right) as compared to VMD. 
    The only theory curves are the dashed
    VMD lines. The full and dash-dotted lines are fits to the data. Figures taken from \cite{Arnaldi:2009aa}.}
  \label{fig:NA60om}
\end{figure}

The pion form factor is determined by the reaction $e^+ e^- \to \pi^+ \pi^-$. It constitutes one of the traditional
successes of the VMD scenario \cite{sakurai}. Figure \ref{fig:klingl} shows how well VMD works for this form factor.
In contrast, VMD drastically fails for the omega transition form factor, extracted from the reaction
$\omega \to \mu^+ \mu^- \pi^0$. This is displayed one the left-hand side of figure \ref{fig:NA60om} where the VMD 
prediction is
compared to data from NA60 and Lepton G \cite{Landsberg:1986fd,Arnaldi:2009aa}.
For the eta transition form factor, extracted from the reaction $\eta \to \mu^+ \mu^- \gamma$, VMD works very well
as can be seen on the right-hand side of figure \ref{fig:NA60om}. 
Also for the pion transition form factor, extracted, e.g., from $e^- \gamma \to e^- \pi^0$, the VMD scenario
works well. Here, a key quantity is the slope of the form factor at vanishing mass of the virtual photon,
\begin{eqnarray}
  \Lambda^{-2} := \left. \frac{dF(s)}{ds} \right\vert_{s=0}  \,.
\end{eqnarray}
The VMD prediction is ($m_V$ denotes the $\rho$-meson mass)
\begin{eqnarray}
  \Lambda_{\rm VMD} = m_V \approx 0.77\, \mbox{GeV}
\end{eqnarray}
which is in excellent agreement with the experimental result \cite{Nakamura:2010zzi}
\begin{eqnarray}
  \Lambda_{\rm PDG}= (0.76 \pm 0.05)\,\mbox{GeV.}
\end{eqnarray}
This quantity concerns the {\it single-virtual} pion transition form factor, 
i.e.\ the electromagnetic transition between a pion and a real photon. Also the {\it double-virtual} 
pion transition form factor is of considerable interest. For example, it provides an important part of the
hadronic light-by-light scattering contribution to the magnetic moment of the muon \cite{Jegerlehner:2009ry}. 
Therefore, it is important to clarify whether the double-virtual pion transition form factor can be reliably
described by VMD (see, e.g., \cite{Borasoy:2003yb,Petri:2010ea,carla-diplom}). 
This issue has not been settled yet from the experimental side. In principle, reactions like
$\pi^0 \to 2 e^+ \, 2e^-$ and $e^- e^+ \to e^- e^+ \, \pi^0$ have to be studied. 
We will come back to this issue below.

\section{Effective field theories}

In the following we will describe first steps towards an effective field theory for vector mesons. One aim
of such an enterprise is the reliable calculation of electromagnetic form factors and other quantities where
vector mesons constitute important degrees of freedom. In that way it should be possible to figure out where or 
under which
conditions VMD works and where it does not. It should be stressed right away that at present this development is in
an infant stage. Though first results are encouraging --- as will be shown below --- it has not been established yet
that a systematic low-energy expansion is feasible and convergent for the energy region where vector mesons are 
important. Below, a Lagrangian for light vector and pseudoscalar mesons will be presented which is inspired by
the ideas of effective field theories. Without an established systematic power counting this Lagrangian might still
be regarded at present as a --- rather successful --- phenomenological tree-level model. However, the ambition of this
enterprise is higher and one-loop calculations are under way to scrutinize the validity of this approach as a
systematic effective field theory. 

The basic idea of an effective field theory is to perform a Taylor expansion in the involved momenta. (For this
general outline we do not specify whether we talk about four- or three-momenta --- see the more detailed discussion
below.)
This is a reasonable approach as long as one sticks to processes with low enough momenta. The crucial question
is what ``low enough'' means quantitatively. We will come back to this question below. The difference 
between an effective field theory and a phenomenological model (like, e.g., VMD) is that for the construction of the 
Lagrangian one considers {\it all} possible terms
which are in agreement with the symmetries of the system.\footnote{For the strong interaction in the sector of
hadrons made out of light quarks this includes parity, charge conjugation and approximate chiral symmetry.}
Then, the basic idea is to classify these possible terms according to their number of derivatives. The reason
simply is that derivatives translate to momenta. Thus, the more derivatives a term in the Lagrangian has, the less
important it is for low-energy processes. The generic feature of an effective field theory is that one can 
systematically improve a calculation by working out the next order in the Taylor expansion. 
In that sense it is
like an ordinary perturbative field theory. The difference is just that one does not expand in powers of the 
coupling constants but in powers of the involved (small!) momenta. Typically, higher powers in the Taylor expansion
involve higher-loop calculations. In contrast, in a phenomenological model one demands from outside to restrict the
calculations, e.g., to tree level or to one-loop level 
without a reason why the neglected contributions should be small. 
An effective field theory is
defined by the most general Lagrangian one can write down together with a power counting 
(and the quantization procedure).
All emerging Feynman diagrams can be classified according to the power counting. Therefore, one knows at which
order in the Taylor expansion which diagrams contribute. 
In contrast, a phenomenological model is defined by a specific Lagrangian together with a ``by-hand''
prescription which Feynman diagrams should be considered.  
In that sense an effective field theory is systematic while a phenomenological model is not. On the other hand,
one cannot always formulate an effective field theory for a given energy region. The point is that an effective
field theory is only feasible, i.e.\ {\it effective}, if the Taylor expansion in the involved momenta 
converges (fast enough). Therefore the quantitatively important question is: What is the {\it dimensionless}
quantity the Taylor expansion (power counting) is based on? 
In the numerator one has the momenta, the so-called ``soft scale''.
But which quantity appears in
the denominator? This ``hard scale'' needs to be significantly larger than the considered momentum region. Only then
the effective field theory can work. This discussion will be continued below when we have specified the respective
Lagrangians. 

In the following we restrict ourselves to mesons built out of the lightest two quark flavors, namely
the pseudoscalar pions, encoded in the flavor matrix
\begin{eqnarray}
  \label{eq:pionfield}
  \Phi = \left(\begin{array}{cc}
        \pi^0 & \sqrt{2}\,\pi^+ \\
        \sqrt{2}\,\pi^- & -\pi^0 
      \end{array}\right)   \,,
\end{eqnarray}
and the rho and omega vector mesons, collected in
\begin{eqnarray}
  \label{eq:vectorfiuelds}
  V_{\mu \nu} = \left(\begin{array}{cc}
        \rho^0+\omega & \sqrt{2}\,\rho^+ \\
        \sqrt{2}\,\rho^- & -\rho^0+\omega 
      \end{array}\right)_{\mu \nu}  \,.
\end{eqnarray}
Strictly speaking, most of the results which will be presented below have been obtained in a framework 
\cite{Lutz:2008km,Leupold:2008bp,Terschluesen:2010ik,carla-diplom}
which included three active
quark flavors --- up, down and strange. There, the octet of pseudoscalar mesons ($\pi$'s, $K$'s, $\eta$) together with
the nonet of vector mesons ($\rho$'s, $K^*$'s, $\omega$, $\varphi$) has been taken into account. 
However, a fully consistent framework had to include the full flavor nonets of pseudoscalar and vector mesons.
This implies that one needs to incorporate also the flavor singlet $\eta$ and its mixing with the octet $\eta$ 
to the physical states $\eta$ and $\eta'$. 
Such an extension is presently under construction \cite{TLL}. 

For the energy region where the light pseudoscalar mesons are the only relevant degrees of freedom the appropriate
effective field theory is well established; it is called 
``chiral perturbation theory'' ($\chi$PT) \cite{Weinberg:1978kz,Gasser:1983yg,Gasser:1984gg,Scherer:2002tk}.
We shall briefly review the respective leading-order chiral Lagrangian for the sector with an even and with an
odd number of pseudoscalar mesons, respectively. Afterwards we shall discuss the inclusion of vector mesons.

In the sector of natural parity --- where one only has terms with an even number of pseudoscalar mesons and where
on the formal level no Levi-Civita tensor appears --- the part of the leading-order $\chi$PT Lagrangian relevant for
the present purpose is given by \cite{Weinberg:1978kz,Gasser:1983yg,Gasser:1984gg,Scherer:2002tk}
\begin{eqnarray}
  \label{eq:LOchipt}
  {\cal L}_{\chi\rm PT} = f^2 \, {\rm tr}(U^\dagger_\mu \, U^\mu) 
      + \frac14 \, f^2 \, m_\pi^2 \, {\rm tr}(U^\dagger \! + \! U)   \,.
\end{eqnarray}
Here, ``tr'' is the flavor trace, $f \approx 90 \,$MeV denotes the pion decay constant and $m_\pi$ the pion mass.
The pion fields are encoded in
\begin{eqnarray}
  \label{eq:LOchipt2}
  U= u^2 = \exp({\rm i}\,{\Phi}/{f}) \,, \qquad U_\mu = \frac12 \, u^\dagger \, (D_\mu U) \, u^\dagger
\end{eqnarray}
with the gauge covariant derivative 
\begin{eqnarray}
  \label{eq:gcder}
  D_\mu U = \partial_\mu U + i e \, [Q,U] \, A_\mu  \,,
\end{eqnarray}
the positron charge $e$,
the quark charge matrix
\begin{eqnarray}
  \label{eq:quarkcharge}
  Q = \left(\begin{array}{cc}
        + \frac23 & 0 \\
        0 & - \frac13 
      \end{array}\right) 
\end{eqnarray}
and the photon field $A_\mu$. We have neglected isospin breaking and external fields other than the electromagnetic
one.
The non-linear appearance of the pion fields (\ref{eq:pionfield})
in (\ref{eq:LOchipt2}) is caused by the spontaneous breaking of 
chiral symmetry. One implication is that already the chirally extended kinetic term, 
$f^2 \, {\rm tr}(U^\dagger_\mu \, U^\mu)$, contains interactions with a fixed interaction strength given by the
pion decay constant. For example, the corresponding four-point interaction is 
$\sim \frac{1}{f^2} \, [\Phi,\partial_\mu \Phi] \, [\Phi,\partial^\mu \Phi]$. 

For reactions of pions with pions and/or photons the tree-level diagrams emerging from the leading-order
Lagrangian provide a decent description --- as long as the reaction energy is small 
enough.\footnote{What ``small enough'' means quantitatively will be specified in a moment.} If one assigns
a generic momentum $p$ to such reactions, then all involved particle energies and three-momenta should roughly be
as large as $p$ (or smaller). As a consequence also the pion mass should be of this order, i.e.\ $m_\pi \sim p$.
The terms in the leading-order Lagrangian (\ref{eq:LOchipt}) are all of order $O(p^2)$: In the first term the
derivatives translate to momenta which are $\sim p$. In the second term one has the square of the pion mass.
It is advantageous to count also the photon field as order $p$. In that way, both terms in the gauge covariant
derivative (\ref{eq:gcder}) have formally the same order.

There can be no terms of order $O(p^0)$ because of the Goldstone-boson nature of the pions. There can be
no terms of order $O(p^1)$ because of Lorentz invariance. Thus, $O(p^2)$ is the leading order and 
it is important to note that (\ref{eq:LOchipt}) is the most general Lagrangian of order $O(p^2)$ which is in agreement
with the symmetries of the strong interaction, i.e., for the case at hand, parity, charge conjugation and approximate chiral symmetry. For the given setup, i.e.\ considering only pions as active degrees of freedom, 
there are no other terms of order $O(p^2)$. Thus, there are only two parameters (``low-energy constants'') 
governing the lowest-order
Lagrangian: the pion decay constant $f$ and the pion mass $m_\pi$. They can be adjusted to data or ideally
deduced from the microscopic theory, QCD. Once these parameters are determined, the Lagrangian has predictive
power for low-energy reactions. In addition, one can show that loop diagrams emerging from
(\ref{eq:LOchipt}) are at least of order $O(p^4)$. Therefore, a tree-level calculation is sufficient at low enough
energies.

If one wants
to describe reactions at somewhat higher energies or the low-energy reactions with larger accuracy, then one has
to work out the $O(p^4)$ contributions. They emerge from one-loop diagrams based on the leading-order Lagrangian
(\ref{eq:LOchipt}) and from tree-level diagrams based on the next-to-leading-order Lagrangian. The latter
contains all terms with four derivatives and/or pion masses. It involves additional low-energy constants which
again needed to be fitted to additional data or deduced from QCD. In principle, one can continue this procedure to
higher powers of $p$. In practice, one limitation comes from the plethora of additional parameters
which come in with the higher-order Lagrangians \cite{Bijnens:1999sh,Ebertshauser:2001nj}. 
This restricts the predictive power of more accurate calculations. 

However, there is an additional problem if one would like to increase the applicability range to higher energies. 
The convergence of the power series in $p$ is limited for, at least, two reasons. 
One limit emerges from the loop diagrams. The power series in $p$ breaks down if the loop diagrams
become as important as the tree diagrams based on the same Lagrangian. If one works that out for chiral
perturbation theory one finds that the scale is set by 
\begin{eqnarray}
\label{eq:4pif}  4\pi \, f \approx 1 \, \mbox{GeV.}
\end{eqnarray}
In the following, this scale emerging from a comparison of tree-level and loop diagrams is called ``intrinsic hard
scale''. 

A second limitation comes from the
fact that only the pions (or, for three flavors, the pions, kaons and eta) are taken into account. Therefore,
the convergence of the series must break down in the energy region where also other mesons become active degrees
of freedom. For example, one might consider the $\rho$ meson. It contributes to pion-pion scattering and 
the corresponding Feynman diagrams contain a vector-meson propagator, $1/(p^2-m_V^2)$, 
cf., e.g., the right-hand side of figure \ref{fig:feynrescatt} below. Again, $p$ denotes the
typical energy/momentum of the considered reaction, and $m_V$ denotes the vector-meson mass. For $p \ll m_V$ one
can expand this vector-meson propagator in a power series of $p^2/m_V^2$. In chiral perturbation theory, where
vector mesons are not considered explicitly, such contributions are encoded in the low-energy 
constants \cite{Ecker:1988te,Ecker:1989yg}. Obviously the expansion of the vector-meson propagator breaks down
if the typical momentum becomes as large as the mass of the vector meson. More generally, the masses of the
not considered degrees of freedom set an energetic limit to the applicability range of an effective field theory.
This range scales with the mass gap between the heaviest considered and the lightest not considered state of the
effective field theory. For chiral perturbation theory it is very advantageous that the pions are much lighter
than all the other mesons because of the Goldstone-boson character of the pions. The scale connected to the
neglected degrees of freedom is called ``external hard scale'' in what follows.

In principle, it is
tempting to extend chiral perturbation theory by including heavier states. In that way one 
could include more processes and it might be possible to push the external hard scale 
to higher energies. There are several aspects to consider:
One has to decide whether one wants to include these additional degrees of freedom as light or as heavy
as compared to the considered momentum scale $p$. For light degrees of freedom all masses, energies and momenta
should be of the order of $p$. To turn the argument around, the masses and momenta define the soft scale $p$. This makes only sense if the soft scale remains significantly smaller than the intrinsic hard scale introduced above.
This provides a limit for the masses of states which could be reasonably considered as ``light''.
In addition, the convergence of the power series requires a significant gap between the heaviest of the
included states and the lightest of the not included states. For chiral perturbation theory the large
gap emerges from the fact that one deals with Goldstone bosons. For the rest of the hadron spectrum the
identification of a sizable gap is not so obvious. This issue will be picked up below.

For additional heavy degrees of freedom a power series in terms of the {\it three}-momenta of the heavy states 
and the 
four-momenta of the light states might be reasonable. Here the masses of the light states and the involved 
three-momenta of all states define the soft scale $p$. The mass(es) of the heavy states are part of the hard scale.
This is the framework of baryon chiral perturbation theory \cite{Scherer:2002tk}.
Here it is crucial that the baryons cannot decay into mesons, in particular not into light mesons. 
Otherwise the scales are intertwined, because the three-momenta of the light decay products inherit the large scale
of the decaying heavy state. This is one of the aspects which makes the inclusion of vector mesons so challenging
(see below).

We have determined {\it quantitatively} what the 
phrase ``an effective field theory works at low enough energies''
means for strict mesonic $\chi$PT. The lowest not considered degree of freedom is the sigma meson with a mass of about
600 MeV \cite{Nakamura:2010zzi}. This external hard scale seems to set the actual limit because it is lower than the intrinsic hard scale (\ref{eq:4pif}) set by the loops. 

What was presented so far is the conservative picture. 
However, there might be some twists to the previous arguments: There are studies which suggest that the sigma meson
is not a quark-antiquark state but rather a two-pion correlation (see, e.g., \cite{Oller:1998zr,Danilkin:2011fz} 
and references
therein). If this is true, then there is no compelling reason to include the sigma as an explicit degree of freedom 
in an effective Lagrangian. An illustrative analogy might be the deuteron which emerges as a bound state from
a calculation based on a Lagrangian which contains only nucleons (and mesons). In a similar way, 
the sigma could appear ``dynamically'' in the pion-pion phase shift. The Lagrangian might not contain the sigma
meson. 
However, such a formalism can only work if the power counting is {\it not} directly 
applied to the
observable quantities like the scattering amplitude. This would be like a {\it truncated} Born series applied to the
problem of a bound state or a resonance. One rather has to sum up the {\it whole} Born series, i.e.\ really solve the
Lippmann-Schwinger equation instead of a perturbative treatment. This translates to the demand to ``unitarize''
$\chi$PT. The power counting should be applied to the determination of scattering kernels and not to scattering
amplitudes. In other words, rescattering processes need to be resummed. 
From the puristic point of view this concept is unsatisfying because there are ambiguities in the
unitarization/resummation procedure. 
On the other hand, unitarized versions of $\chi$PT have produced very promising results 
in the last few years \cite{GomezNicola:2001as,Oller:1998zr,Danilkin:2011fz}. In spite of the mentioned ambiguities the common feature of
such approaches is the resummation of specific classes of diagrams (the two-particle reducible diagrams which
correspond to rescattering). 
In such a framework the intrinsic hard scale needs to be re-evaluated. One needs to compare the size of
two-particle irreducible loops and tree-level terms. Such an analysis has not been carried out yet. However,
the successes of unitarized versions of $\chi$PT up to energies of 1 GeV and beyond suggest that the actual
intrinsic hard scale is higher up than (\ref{eq:4pif}) once the two-particle reducible diagrams are properly resummed.

We have seen that it might be possible to circumvent both limitations of strict $\chi$PT, the sigma mass and the 
scale (\ref{eq:4pif}). The next external hard scale is provided by the vector mesons. From all we know about QCD
at present, it is highly unlikely that the vector mesons are two- or three-pion correlations. They are
dominantly quark-antiquark states (see, e.g.\ \cite{Leupold:2009nv} and references therein). Therefore, they
need to be included as explicit degrees of freedom in an effective Lagrangian which is supposed to work in the
region of low-lying mesonic resonances. If this is possible at all is a matter of active research. Further discussions 
and first steps towards this goal are presented in the next section.
Before, however, the leading-order chiral Lagrangian for the sector of an odd number of pions will be briefly
discussed. 

In the sector of anomalous parity --- where on the formal level a Levi-Civita tensor $\varepsilon_{\mu\nu\alpha\beta}$ 
is involved ---
the leading-order contribution to the effective field theory for Goldstone bosons is given by the Wess-Zumino-Witten
action \cite{Wess:1971yu,Witten:1983tw}. It is fully determined by the chiral anomaly and free of any undetermined
low-energy parameters. It contributes at order $O(p^4)$. For the present purpose the only relevant part is the
$\pi^0 \gamma \gamma$ term given by
\begin{eqnarray}
  {\cal L}_{\rm WZW} = \frac{3}{32\pi^2} \, \frac{e^2}{{f}} \, 
  \epsilon_{\mu\nu\alpha\beta} \, F^{\mu\nu} \, F^{\alpha\beta} \, {\rm tr}(Q^2 \, \Phi)
  \label{eq:WZW}
\end{eqnarray}
with the electromagnetic field strength $F_{\mu\nu} = \partial_\mu A_\nu - \partial_\nu A_\mu $. 
This parameter-free Lagrangian provides an excellent description of the decay of the neutral pion into two real
photons \cite{Scherer:2002tk}. It will constitute one contribution to the pion transition form factor, see
section \ref{sec:results} below.

\section[Vector mesons]{The vector-meson Lagrangian and its parameters}

In the last section the effective field theory for Goldstone bosons, $\chi$PT, has been introduced. It has also
been used to illustrate the general ideas of an effective field theory. For the formulation of an effective
field theory for the energy region of vector mesons one has to consider the following questions:
Should the vector mesons been introduced as light or heavy degrees of freedom? If light, which other
mesons limit the applicability range, i.e.\ is there a large enough gap between the considered and not considered
states? If light, where is the intrinsic hard scale? If heavy, how to deal with the fact that vector mesons can decay
into light degrees of freedom (pions, photons)?

In the following, only preliminary answers can be provided which need to be further scrutinized in the future.
However, one remarkable feature is that basically {\it the same Lagrangian} emerges (at the respective leading
order), no matter whether one treats the vector mesons as light or as heavy degrees of freedom! This gives some
credit to the proposed Lagrangian and might to some extent explain its phenomenological success even if a 
full-fledged effective field theory has not been established yet (or maybe even cannot be established). 

The inclusion of vector mesons as light degrees of freedom has been proposed in \cite{Lutz:2008km} 
and further explored in \cite{Leupold:2008bp,Terschluesen:2010ik,Danilkin:2011fz}. Based on the hadrogenesis 
conjecture \cite{Lutz:2001yb,Lutz:2001mi,Lutz:2003fm,Lutz:2008km} it has been argued that the other mesons in the energy range close to the vector mesons
are dynamically generated from the coupled-channel interactions of pseudoscalar and vector mesons. 
Therefore, such states need not be considered explicitly in the effective Lagrangian. This implies that 
there is a sizable gap between the energy region of the lowest-lying vector and pseudoscalar mesons and the not 
considered other (quark-antiquark) states. 

Whether the intrinsic hard scale is significantly larger than the vector-meson masses needs to be explored in loop
calculations. It should be stressed again that exact unitarity, i.e.\ the resummation of two-particle 
reducible diagrams, is mandatory for an effective field theory which operates in the region of hadronic resonances. 
As already mentioned, the successes of unitarized versions of $\chi$PT suggest that the intrinsic hard scale could
be significantly larger than 1 GeV. This would provide some room where an effective field theory of light vector
mesons can be applicable. 

The inclusion of vector mesons as strict heavy degrees of freedom has been pioneered in \cite{Jenkins:1995vb}. The
focus has been set to reactions where a constant number of vector mesons remained throughout the whole respective
process. In that way, the problem how to deal with decaying vector mesons has been avoided by hand. In the
present work a more pragmatic way is suggested: Vector mesons are neither truly heavy, nor truly light.
Their mass is close to the intrinsic hard scale (\ref{eq:4pif}) of (strict) $\chi$PT. In that sense they are heavy.
On the other hand, if vector mesons decay, the momenta of the decay products are still small compared to the scale
(\ref{eq:4pif}). Slightly exaggerating, one might say that the vector-meson mass, $m_V$, is still
heavy while $m_V/2$ is already soft. This suggests to discuss separately the Lagrangian which includes the terms
with one vector-meson field and the Lagrangian which contains the terms with two vector-meson fields.\footnote{More
vector-meson fields are not needed at tree level for the processes of interest.} The first Lagrangian concerns
the decays of vector mesons while the second Lagrangian concerns the propagation of vector mesons while absorbing or
emitting light particles. 
Whether this pragmatic point of view is feasible, i.e.\ provides a systematic approach for loop calculations,
needs to be seen in the future.  

The leading-order Lagrangian collecting all terms with one vector-meson field is given by
\begin{eqnarray}
  {\cal L}_1 =   -\frac{{\rm i}}{4} \, {h_P} \, m_V \, {\rm tr}({V_{\mu\nu}} \, [U^\mu , U^\nu]) 
  - \frac18 \, {e_V} \, m_V \, {\rm tr}({V^{\mu\nu}} \, Q) \, F_{\mu\nu}    \,.
  \label{eq:vectorL1}
\end{eqnarray}
The vector mesons are represented by anti-symmetric tensor fields, $V_{\mu\nu}$ \cite{Ecker:1988te,Lutz:2008km}. 
The phrase ``leading order'' has the following meaning: If one considers the vector mesons as semi-heavy in the sense
that their mass is large (part of the hard scale) but the momenta of the decay products are already small $\sim p$, 
then one might count all derivatives (and the photon field encoded in the field strength $F_{\mu\nu}$)
as soft in (\ref{eq:vectorL1}). The terms appearing in (\ref{eq:vectorL1}) are the only ones which have 
one vector-meson field and appear at order $O(p^2)$. Here, it is crucial that the vector mesons are
represented by anti-symmetric tensor fields and not, e.g., by vector fields, $V_\mu$. 
In the latter case one would need
more derivatives to construct the corresponding terms \cite{Ecker:1989yg}, e.g.\ one would have 
${\rm tr}({\partial_\mu} V_\nu \, [{U^\mu} , {U^\nu}]) \sim O(p^3)$ instead of the $h_P$ term in (\ref{eq:vectorL1}). 
It is important to stress that this statement applies to vector-meson fields which transform as ordinary matter
fields with respect to chiral transformations. The treatment of vector mesons as 
gauge bosons \cite{Harada:2003jx} is not
considered in the present work. The use of the anti-symmetric tensor representation is singled out by the fact
that only here $O(p^2)$ terms are possible for structures with only one vector meson. In any other representation
such structures would be at least $O(p^3)$. 

If one considers the vector mesons as light degrees of freedom \cite{Lutz:2008km}, the counting is
the same, all derivatives are of order $p$. The complete Lagrangian is given in (\ref{eq:vectorL1}) and is of order
$O(p^2)$. 

The two
parameters $h_P$ and $e_V$ can be determined from the decays $\rho \to 2\pi$ and $\rho \to e^+ e^-$ 
\cite{Lutz:2008km}. The same results, but with higher accuracy, are obtained from a fit to the 
pion form factor \cite{Leupold:2009nv}. 

For structures with two vector mesons the pertinent Lagrangian is
\begin{eqnarray}
  {\cal L}_2 & = & -\frac14 \, {\rm tr}(D_\mu {V^{\mu\alpha}} \, D^\nu {V_{\nu\alpha}}) 
  + \frac18 \, m_V^2 \, {\rm tr}({V_{\mu\nu}} \, {V^{\mu\nu}})  
%\nonumber \\  && {} %
+ \frac{{\rm i}}{8} \, {h_A} \, \varepsilon_{\mu\nu\alpha\beta} \, 
  {\rm tr}\left(\{ V^{\mu\nu} ,D_\lambda V^{\lambda\alpha} \} \, U^\beta\right)  \,.
  \label{eq:vectorL2}
\end{eqnarray}
Terms with more than one flavor trace have been neglected on account of the 
OZI rule \cite{Gasser:1984gg,Ecker:1988te}. As a consequence
the masses of rho and omega meson agree in this approach. Both are given by $m_V$. 

If the vector
mesons are considered as (semi-)heavy, then (\ref{eq:vectorL2}) is the complete leading-order $O(p)$
Lagrangian. It resembles the case of the nucleon $\chi$PT Lagrangian \cite{Scherer:2002tk}. 
The $h_A$ interaction term translates to the well-known $g_A$ term of pion-nucleon interaction. 

If the vector mesons are considered as light degrees of freedom \cite{Lutz:2008km}, 
one counts all derivatives appearing in
(\ref{eq:vectorL2}). The Lagrangian is of order $O(p^2)$. There are several more terms of this order. However, 
only one contributes to the processes of interest here. This term involves the square of the pion mass and
is numerically rather unimportant \cite{Lutz:2008km,Terschluesen:2010ik}. 

\begin{figure}[h] 
  \centering
  \includegraphics[keepaspectratio,width=0.3\textwidth]{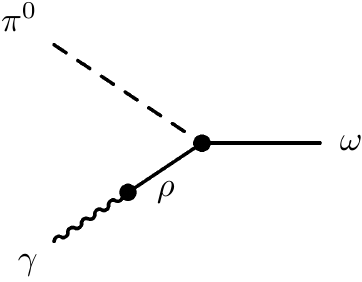}  
  \caption{Two-step decay of an omega meson into a pion and a real photon.}
  \label{fig:omraddec} 
\end{figure}      
The single free parameter $h_A$ can be fitted to the decay of an omega meson into a pion and a real photon.
The corresponding diagram is shown in figure \ref{fig:omraddec}. 
Note that the process proceeds in two steps
which involves the coupling constants $h_A$ for the $\omega$ to $\rho\pi$ decay and the constant $e_V$
from (\ref{eq:vectorL1}) for the direct coupling of the virtual $\rho$ to the photon. Since $e_V$ has been
determined from another reaction ($\rho \to e^+ e^-$), the process $\omega \to \pi^0 + \gamma$ can be used to
fix $h_A$ \cite{Lutz:2008km,Terschluesen:2010ik}. 
A direct interaction term for $\omega \to \pi^0 + \gamma$ would involve more
than two derivatives, i.e.\ it is of higher power than the terms of (\ref{eq:vectorL1}).

To summarize this section: No matter whether one treats vector mesons as semi-heavy or light degrees of freedom,
one ends up with the Lagrangians presented in (\ref{eq:vectorL1}), (\ref{eq:vectorL2}). This is rather encouraging
in view of the fact that the vector-meson mass is in between the hard and the soft scales of $\chi$PT.

In the following,
these Lagrangians are used together with the Lagrangians of $\chi$PT as given in (\ref{eq:LOchipt}) and 
(\ref{eq:WZW}). Their consequences are worked out at tree level or, if needed, including rescattering
effects. Power counting issues will not be discussed anymore but deferred to future work.

\section[Form factors]{Electromagnetic form factors}
\label{sec:results}

In the Lagrangians (\ref{eq:vectorL1}), (\ref{eq:vectorL2}) there are three free parameters. As already
discussed they can be determined from two-body decays: $h_p$ from $\rho \to 2\pi$, 
$\vert e_V \vert$ from $\rho \to e^+e^-$
and $\vert h_A \cdot e_V \vert$ from $\omega \to \pi^0 \gamma$. 
The numerical results are \cite{Lutz:2008km,Leupold:2009nv,Terschluesen:2010ik}
\begin{eqnarray}	
  \vert e_V \vert \approx 0.22 \,, \qquad h_P \approx 0.30 \,, \qquad  \vert h_A \vert \approx 2.1  \,.
  \label{eq:num} 
\end{eqnarray}
There is a freedom to choose the sign of one coupling constant which multiplies an odd number of vector-mesons fields
and the sign of one coupling constant which multiplies an odd number of pseudoscalar fields. This freedom has been
used to specify the overall sign in the Wess-Zumino-Witten term (\ref{eq:WZW}) and the sign of the parameter $h_P$.
To pin down the signs of the parameters $e_V$ and $h_A$ one needs more complex reactions. In other words:
The sign of $h_P$ is pure convention, but the sign of $e_V$ relative to $h_P$ is a physics issue. The same applies
to the sign of $h_A$ relative to the Wess-Zumino-Witten term (\ref{eq:WZW}).

Having fixed the parameter values (at least their absolute sizes), no further parameters are needed to determine
\begin{itemize}
\item the pion form factor, contained in the reaction $e^+ e^- \to \pi^+ \pi^-$,
\item the omega transition form factor, $\omega \to \pi^0 \, l^+ l^-$,
\item the single-virtual pion transition form factor, $\pi^0 \to \gamma \, e^+ e^-$,
\item the double-virtual pion transition form factor, $\pi^0 \to e^+ e^- \, e^+ e^-$.
\end{itemize}
It is important to note that at no stage any VMD assumption has entered. It might come out
as a result of the calculations, but it is not put in by hand.

\begin{figure}[h] 
  \includegraphics[keepaspectratio,width=0.45\textwidth]{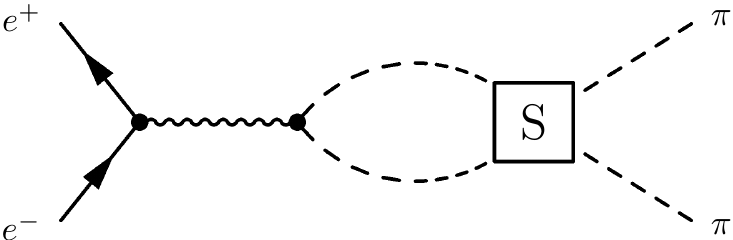} \hfill 
  \includegraphics[keepaspectratio,width=0.45\textwidth]{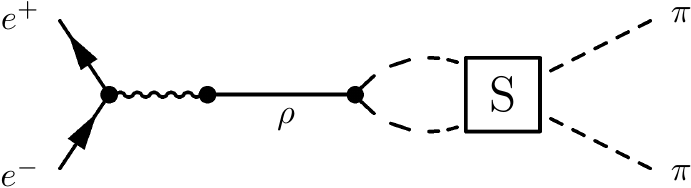}
  \caption{The two possible processes which lead from the initial state $e^+ e^-$ to the final state $ \pi^+ \pi^-$.
  The box denoted by ``S'' signals the rescattering of the pions. Figure taken from \cite{Leupold:2009nv}.}
  \label{fig:feynpionFF}
\end{figure}      
The process $e^+ e^- \to \pi^+ \pi^-$ can take place via a direct coupling of the pions to the photon 
or via a vector meson. This is depicted in figure \ref{fig:feynpionFF}. 
In the first case, the pions
couple via their electric charge. This is described by the Lagrangian (\ref{eq:LOchipt}). In the second
case the Lagrangian (\ref{eq:vectorL1}) provides the coupling of the virtual photon to the rho meson $\sim e_V$
and the coupling of the rho meson to the pions $\sim h_P$. 

\begin{figure} 
  \includegraphics[keepaspectratio,width=0.4\textwidth]{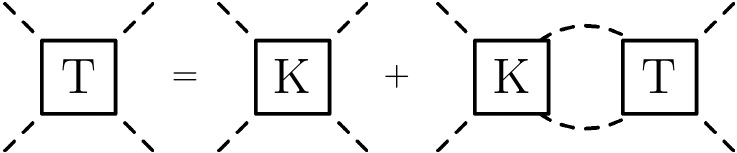} \hfill 
  \includegraphics[keepaspectratio,width=0.45\textwidth]{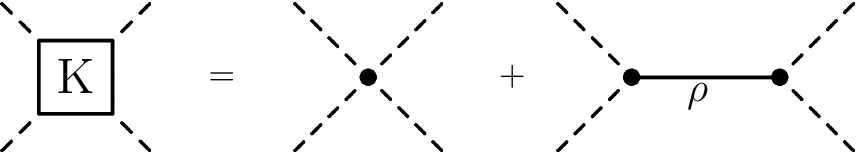}  
  \caption{Bethe-Salpeter equation for the rescattering of pions in the $p$-wave channel. The $T$ matrix of
  pion-pion scattering is obtained from the scattering kernel denoted by $K$. 
  Figure taken from \cite{Leupold:2009nv}.}
  \label{fig:feynrescatt}
\end{figure}

A reliable description of the reaction $e^+ e^- \to \pi^+ \pi^-$ and the pion form factor requires a 
resummation of the rescattering processes of the pions. In \cite{Leupold:2009nv} this has been achieved 
by a Bethe-Salpeter equation. Schematically this is depicted in figure \ref{fig:feynrescatt}. 
The scattering kernel consists of two parts: The direct contact interaction between the pions is obtained
from (\ref{eq:LOchipt}). The interaction between the pions and the rho meson comes again from the $h_P$ term of
(\ref{eq:vectorL1}). 

Concerning the comparison to VMD it is illuminating to discuss first the tree-level result for the pion form factor.
Afterwards the full calculation including the rescattering process will be compared to the data. 

The tree-level result is obtained from the diagrams of figure \ref{fig:feynpionFF} by putting $S$ to 1, i.e.\ by dropping $S$ in the figure. It is common practice to normalize the pion form factor to the direct term
(first term in figure \ref{fig:feynpionFF} when putting $S$ to 1). 
In that way, the pion form factor becomes unity at the photon point.
The tree-level pion form factor, as obtained from the Lagrangians (\ref{eq:LOchipt}) and (\ref{eq:vectorL1}) is
given by
\begin{eqnarray}
  F_\pi(s) = 1 + \frac{ e_V \, h_P \, m_V^2}{16 e \, f^2} \, \frac{s}{m_V^2-s}   \,.
\label{eq:pionFFcalc}
\end{eqnarray}
The invariant mass of the virtual photon is denoted by $s$. 
The VMD prediction is
\begin{eqnarray}
   F^{\rm VMD}_\pi(s) = \frac{m_V^2}{m_V^2-s} \,.
\label{eq:pionFFcalcVMD}
\end{eqnarray}
The two formulae would analytically agree for $e_V \, h_P = 16 e \, f^2/m_V^2 \approx {0.065}$. With the
values (\ref{eq:num}) one obtains $\vert e_V \vert \, h_P \approx  {0.066}$ which is obviously very close, provided
one chooses a positive sign for $e_V$. Then
a cancellation takes place: The two terms from the $\chi$PT Lagrangian (\ref{eq:LOchipt}) 
and the vector-meson Lagrangian (\ref{eq:vectorL1}), respectively, conspire such that the final result is 
close to VMD. 

\begin{figure} 
  \centering
  \includegraphics[keepaspectratio,width=0.5\textwidth]{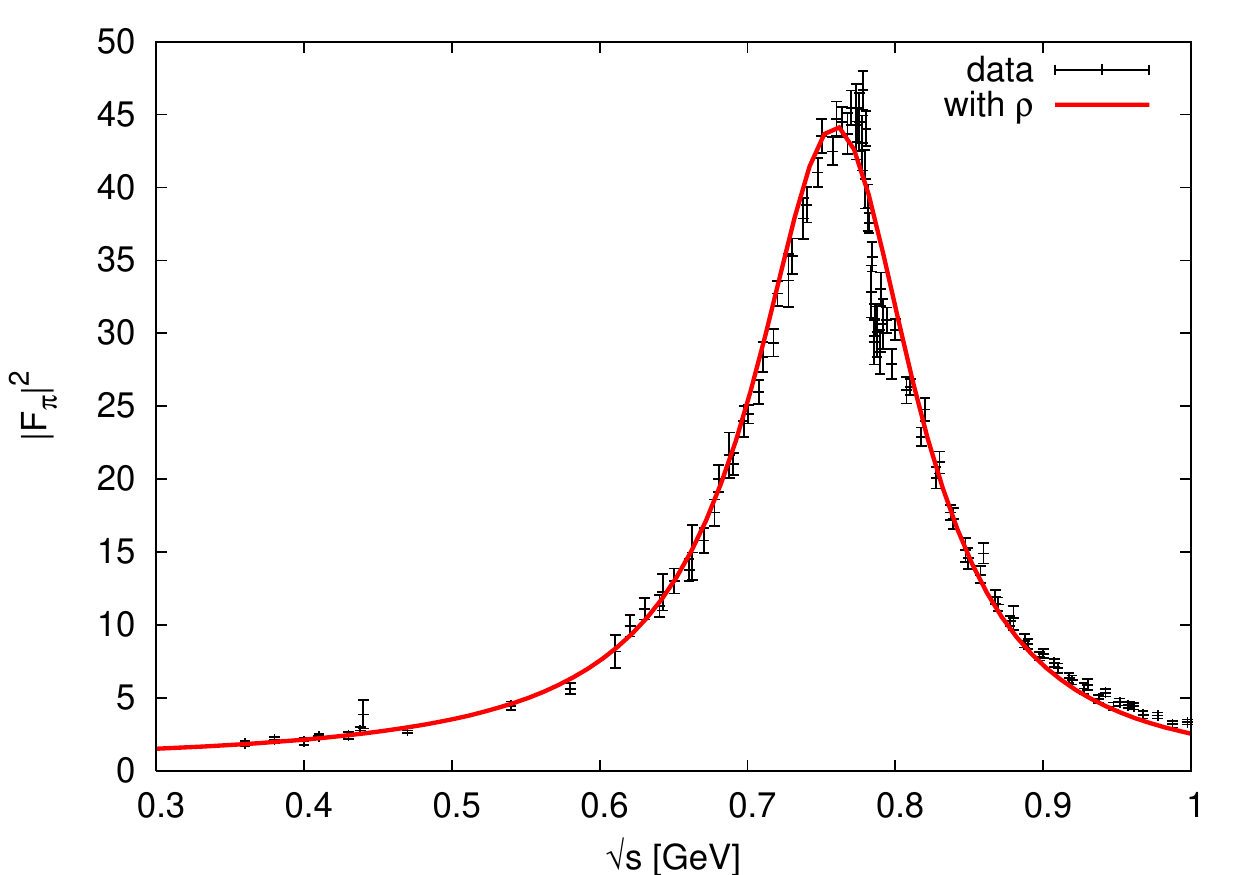}  
  \caption{The pion form factor as compared to data \cite{Leupold:2009nv}.}
  \label{fig:withrho}
\end{figure}

The full result based on the Lagrangians (\ref{eq:LOchipt}) and (\ref{eq:vectorL1}),
but now including the rescattering of pions is shown in figure \ref{fig:withrho}. Obviously one obtains an
excellent result. Note that isospin breaking leads to $\rho$-$\omega$ mixing. This has not been included. Therefore,
the sharp omega peak seen in the data cannot be reproduced. 

\begin{figure} 
  \centering
  \includegraphics[keepaspectratio,width=0.4\textwidth]{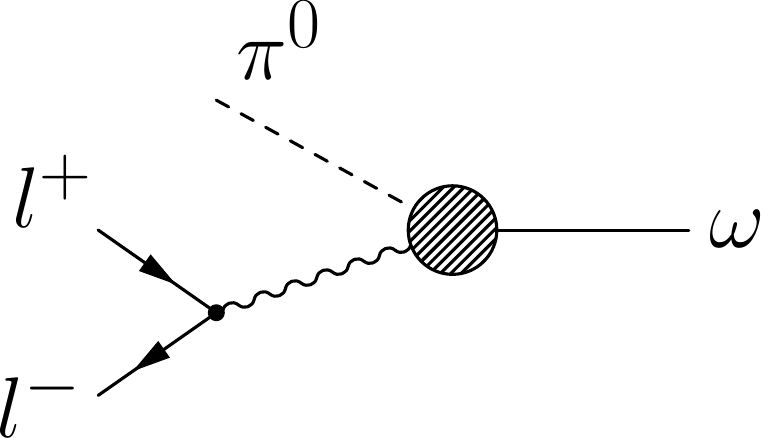}
  \caption{Generic picture for the omega-to-pion transition form factor. $l$ denotes a lepton.}
  \label{fig:omegadalgen}
\end{figure}

\begin{figure}
  \centering 
  \includegraphics[keepaspectratio,width=0.4\textwidth]{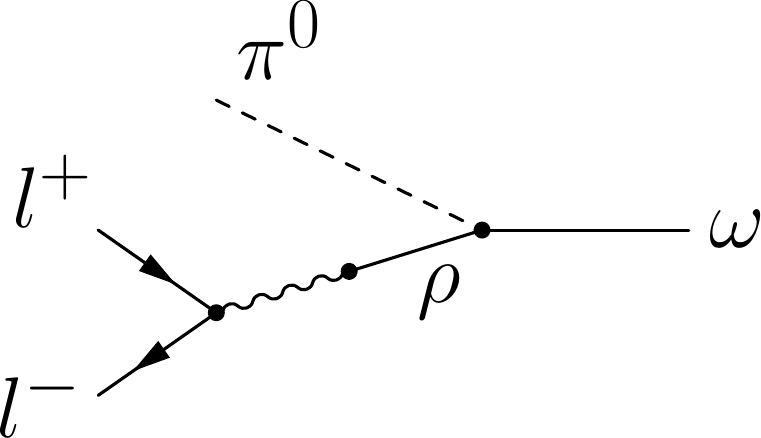}  
  \caption{The omega-to-pion transition form factor in the present approach. The coupling $\rho\omega\pi$ is given
  by $h_A$ and the direct coupling of the rho meson to the photon by $e_V$.}
\label{fig:omegadalLO}
\end{figure}

The next quantity which will be considered is the electromagnetic form factor of the transition between an
omega meson and a pion, see figure \ref{fig:omegadalgen}. 
The form factor is again normalized to the photon point. Thus, it
parametrizes the deviation from a structureless decay.
In the present approach the process is described by the diagram shown in figure \ref{fig:omegadalLO}. 
In principle, the intermediate rho meson could decay into two pions. Thus the rescattering of these pions
which causes the width of the rho meson should be considered. It has been checked, however, that the decay
process $\omega \to \pi^0 l^+ l^-$ with the lepton $l=e,\mu$ is not sensitive to the pion rescattering effects.
Therefore, only the tree-level result will be presented in the following. The case would be different
for the process $e^+ e^- \to \omega \pi^0$ which probes the transition form factor at larger invariant masses. 
There, a proper description would require a coupled-channel calculation at least with the channels $\pi^+\pi^-$
and $\omega \pi^0$, if not $K \bar K$. This is beyond the scope of the present work.

\begin{figure} 
  \centering
  \includegraphics[keepaspectratio,width=0.5\textwidth]{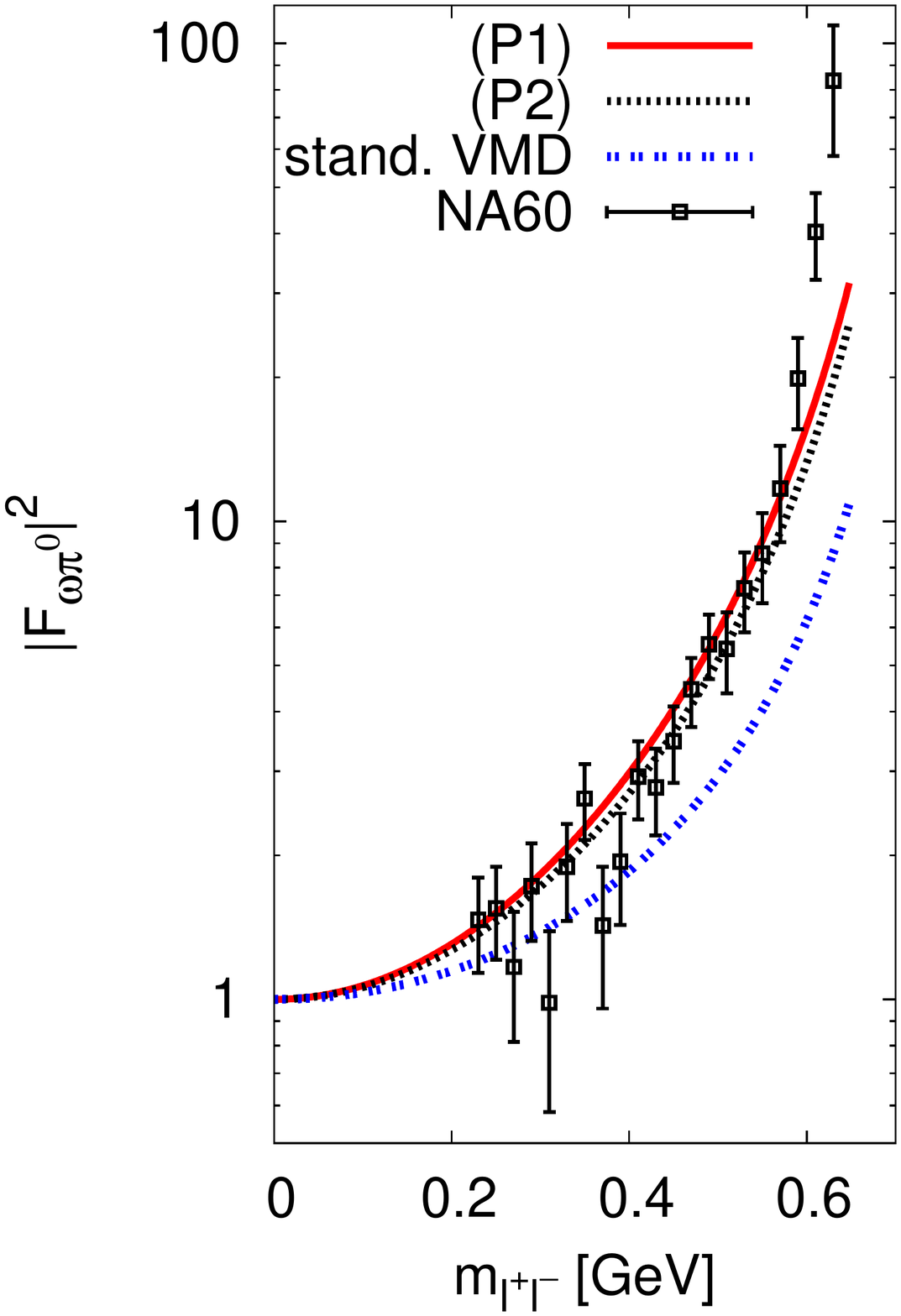}
\caption{The omega to pion transition form factor. The present calculations are depicted by the full (red) line.
  Figure taken from \cite{Terschluesen:2010ik}.}
\label{fig:FFomega}
\end{figure}

The process shown in figure \ref{fig:omegadalLO} is proportional to the product $h_A \cdot e_V$. For the
form factor, however, this overall factor drops out because the form factor is normalized to the photon point.
Therefore, the Lagrangians (\ref{eq:vectorL1}), (\ref{eq:vectorL2}) provide a parameter independent prediction
for the omega transition form factor:
\begin{eqnarray}
  F_{\omega\pi}(s) = \frac{m_V^2+s}{m_V^2-s}    \,.
\label{eq:ompiour}
\end{eqnarray}
This deviates significantly from VMD:
\begin{eqnarray}
  F^{\rm VMD}_{\omega\pi}(s) = \frac{m_V^2}{m_V^2-s}    \,.
\label{eq:ompiVMD}
\end{eqnarray}
Both form factors are compared to data from NA60 \cite{Arnaldi:2009aa} in figure \ref{fig:FFomega}.
Obviously, the NA60 data are much better described by the present approach (\ref{eq:ompiour}) than by the VMD
formula (\ref{eq:ompiVMD}). A follow-up experiment
using dielectrons instead of dimuons is planned by the WASA at COSY collaboration. 

\begin{figure}[h] 
  \centering
  \includegraphics[keepaspectratio,width=0.4\textwidth]{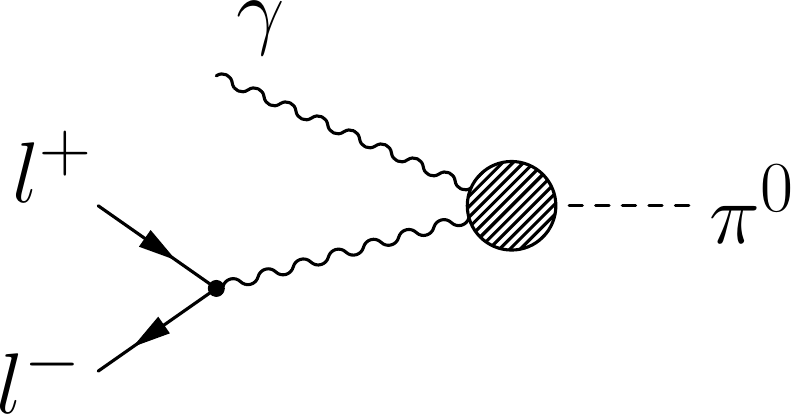}
  \caption{Generic picture for the pion-to-photon transition form factor. $l$ denotes a lepton.}
  \label{fig:piFFgen}
\end{figure}
\begin{figure} 
  \centering
  \includegraphics[keepaspectratio,width=0.25\textwidth]{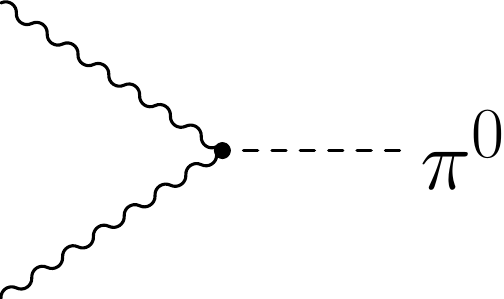}
  \hspace*{5em}
  \includegraphics[keepaspectratio,width=0.25\textwidth]{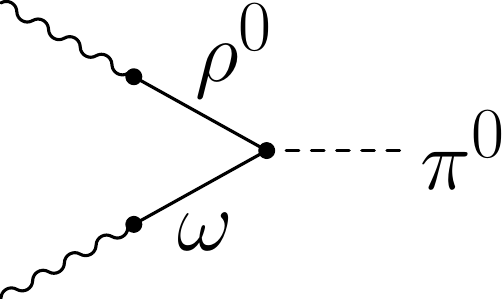}
  \caption{{\it Left:} The Wess-Zumino-Witten action contains a point interaction for one pion and two 
    (real or virtual) photons (wavy lines).
    {\it Right:} The vector-meson Lagrangians (\protect\ref{eq:vectorL1}), (\protect\ref{eq:vectorL2})
    induce an additional interaction between a pion and two photons.}
  \label{fig:piFFWZW}
\end{figure}

      
Next, the pion transition form factor is considered; at first, the single-virtual form factor which corresponds
to the process $\pi^0 \to \gamma \, e^+ e^-$. It is depicted in figure \ref{fig:piFFgen}. 
The present approach has two contributions, 
shown in figure \ref{fig:piFFWZW}, one from the
Wess-Zumino-Witten term (\ref{eq:WZW}) and one from the vector-meson 
Lagrangians (\ref{eq:vectorL1}), (\ref{eq:vectorL2}).
Note that the latter contribution vanishes if both photons are real, i.e.\ for $\pi^0 \to 2\gamma$. Thus, the
excellent description of the latter process by the Wess-Zumino-Witten Lagrangian is not spoiled. 
The pion transition form factor, normalized to the photon point, i.e.\ normalized to the Wess-Zumino-Witten
contribution, is given by
\begin{eqnarray}
  F_{\pi\gamma}(s) = {1} + {\frac{\pi^2 \, h_A \, e_V^2}{12 e^2} \, \frac{s}{m_V^2-s} }  \,.
  \label{eq:piontransformula}
\end{eqnarray}
The corresponding VMD formula is 
\begin{eqnarray}
  F^{\rm VMD}_{\pi\gamma}(s) = \frac{m_V^2}{m_V^2-s}   \,.
  \label{eq:piontransformulaVMD}
\end{eqnarray}
Note that the masses of rho and omega meson are not distinguished here. Of course, there are kinematical situations
where the tree-level results are not enough. For the comparison of the present approach to the VMD result, however,
it is most transparent to look at the tree-level formulae.

The formulae (\ref{eq:piontransformula}) and (\ref{eq:piontransformulaVMD}) would agree analytically for
$h_A \, e_V^2 = 12 e^2/\pi^2 \approx {0.11}$. Using the numerical values from (\ref{eq:num}) one obtains 
$\vert h_A \vert \, e_V^2 \approx  {0.10}$. 
Thus, the formula (\ref{eq:piontransformula}) is again numerically close to the 
VMD result, provided one uses a positive sign for $h_A$. 
Another cancellation has taken place, now between the Wess-Zumino-Witten term and the vector-meson
contribution. As already mentioned in the introduction, the VMD prediction is close to the experimental result.
Therefore, the present approach agrees also with the experimental result for the single-virtual pion transition
form factor. 

%\begin{figure} 
%  \centering
%  \includegraphics[keepaspectratio,width=0.5\textwidth]{Ff-eta-gamma-gemischt}    
%  \caption{.}
%  \label{fig:FFeta}
%\end{figure}      

However, the picture changes for the {\it double-virtual} transition form factor: 
It is given by
\begin{eqnarray}
  F(s_1,s_2) & = & 1 + \frac{\pi^2 \, h_A \, e_V^2}{12 e^2} \, \frac{m_V^2 \, (s_1+s_2)}{(m_V^2-s_1) \, (m_V^2-s_2)} 
  \nonumber \\
  & \approx & 1 + \frac{m_V^2 \,(s_1+s_2)}{(m_V^2-s_1) \, (m_V^2-s_2)}  
  \nonumber \\
  & = & 1 \underbrace{{}-\frac{m_V^2}{m_V^2-s_1} - \frac{m_V^2}{m_V^2-s_2}}_{\mbox{``half'' VMD}} + 2
  \underbrace{\frac{m_V^4}{(m_V^2-s_1) \, (m_V^2-s_2)} }_{\mbox{VMD type}}  
  \label{eq:final}
\end{eqnarray}
while the VMD formula is simply
\begin{eqnarray}
  F_{\rm VMD}(s_1,s_2) = \frac{m_V^4}{(m_V^2-s_1) \, (m_V^2-s_2)}  \,.  
  \label{eq:finalVMD}
\end{eqnarray}
The two virtualities of the photons are denoted by $s_1$ and $s_2$.
Obviously, (\ref{eq:final}) is different from VMD for $s_1,s_2 \neq 0$. 
Future experimental data for the double-virtual transition form factor
are crucial to distinguish between the present approach and the VMD scenario. As already mentioned in the introduction
such data are also an important input for the calculation of the hadronic contribution to the gyromagnetic 
ratio of the muon.

\section[Summary]{Further discussion, summary and outlook}

The respective lowest-order Lagrangians of chiral perturbation theory for the sectors with an even and with an
odd number of pions have been combined with a Lagrangian for vector mesons as given in (\ref{eq:vectorL1}) and 
(\ref{eq:vectorL2}). Also these ingredients have been motivated by effective field theory ideas and different options
and open problems concerning the proper power counting in the energy region of vector mesons have been discussed.
Though several issues are unsettled, it is appealing that the constructed vector-meson Lagrangian can be motivated
in two different way, treating the vector mesons as light or as (semi-)heavy degrees of freedom. 

The developed formalism can be used to calculate the following quantities:
\begin{itemize}
\item All three parameters of the vector-meson Lagrangian are fitted to the two-body decays $\rho \to 2 \pi$, 
  $\rho/\omega \to l^+ l^-$ ($l = \,$lepton) and $\omega \to \pi^0 \gamma$ \cite{Lutz:2008km}.
\item The reaction $e^+ e^- \to \pi^+ \pi^-$ and the corresponding pion form factor is in good agreement with the
  data. Intimately connected to this is also a good description of the $p$-wave 
  pion-pion scattering phase shift (not shown here) \cite{Leupold:2009nv}.
\item The decay $\omega \to \pi^0 \, l^+ l^-$ and the corresponding omega transition form factor is much better described
  than with the vector-meson dominance (VMD) model \cite{Terschluesen:2010ik}. 
\item The single-virtual pion transition form factor, as, e.g., measured in the decay $\pi^0 \to \gamma \, e^+ e^-$,
  turns out to be close to the VMD result and therefore also close to the data.
\item A prediction is provided for the double-virtual pion transition form factor which disagrees with the VMD
  prediction. 
\item The decay rate of $\omega \to \pi^+ \pi^-\pi^0$ is reproduced very well (not shown here) \cite{Leupold:2008bp}. 
\item The reaction $e^+ e^- \to \pi^+ \pi^-\pi^0$ is presently investigated \cite{bruno-master}. Like for the
  pion transition form factor there is an interplay between the Wess-Zumino-Witten term and the vector-meson 
  contribution. 
\end{itemize}

It is worth to summarize the comparison to the VMD model: Where the experimental data are well described by VMD
(pion form factor and single-virtual pion transition form factor),
the present approach contains two terms, one from chiral perturbation theory and one including vector mesons.
These two terms conspire such that cancellations occur which bring the results very close to the VMD results. 
Note that this has not been adjusted
by hand. The coupling constants of the vector-meson Lagrangian can been determined independently
using the two-body decays, not the form factors. For the pion form factor the mentioned cancellation might be
related to the KSFR relation \cite{Ecker:1989yg}. On the other hand, the other cancellation which takes place
for the (single-virtual) pion transition form factor has nothing to do with KSFR. In addition, the KSFR
relation does not follow from chiral symmetry alone. On way to derive it is to use in addition
the high-energy behavior of the pion form factor \cite{Ecker:1989yg}. It is not clear whether high-energy
constraints should be applied to a low-energy theory. Obviously, the observed cancellations are very interesting,
but their microscopic reason remains an open issue at present. 

In any case it is appealing to have a formalism which
is flexible enough to reproduce VMD results but can also go beyond it.
Indeed, for the omega transition form factor the VMD prediction drastically deviates from the data. Here, the
present approach provides only one contribution, not two which could potentially show a cancellation effect. 
{\it Qualitatively} resembling the VMD model, the present approach provides a parameter independent prediction for
the omega transition form factor. However, it is {\it quantitatively} different from VMD, namely
$(m_V^2 +s)/(m_V^2-s)$ instead of $m_V^2/(m_V^2-s)$. The former expression provides a much better description
of the experimental data. 

A challenge for the future is to establish a systematic power counting for the energy region of vector mesons.
Here calculations beyond leading order, including in particular loop calculations, are an important issue. 
A second line of development is the extension to three flavors with the inclusion of the $\eta'$. 
Finally, a better (microscopic?) understanding of the observed cancellations would be desirable 
to further appreciate the
successes of VMD and to see its limitations.

{\bf Acknowledgments:} S.L. thanks M.F.M.\ Lutz for collaboration and discussions. He also acknowledges stimulating discussions
with A.\ Kupsc. Finally he wants to thank the organizers of Bormio 2012 for their excellent work.

\end{document}